\documentclass[pre,twocolumn,showpacs,amsfonts,amsmath,floatfix]{revtex4}

\usepackage{graphicx}
\usepackage{epstopdf}
\usepackage{amssymb}
\usepackage{amsmath}
\usepackage{xspace}
\usepackage{bbold}

\newcommand{\ket}[1]{\ensuremath{|#1\rangle}\xspace}
\newcommand{\bra}[1]{\ensuremath{\langle #1|}\xspace}
\newcommand{\ps}[2]{\ensuremath{\langle #1|#2\rangle}\xspace}
\renewcommand{\vec}[1]{{\mathbf #1}}

\begin{document}

\title{Non-Hermitian Euclidean random matrix theory}

\author{A. Goetschy}
\affiliation{Univ. Grenoble 1/CNRS, LPMMC UMR 5493, Maison des Magist\`{e}res, 38042 Grenoble, France}
\author{S.E. Skipetrov}
\affiliation{Univ. Grenoble 1/CNRS, LPMMC UMR 5493, Maison des Magist\`{e}res, 38042 Grenoble, France}

\begin{abstract}
We develop a theory for the eigenvalue density of arbitrary non-Hermitian Euclidean matrices. Closed equations for the resolvent and the eigenvector correlator are derived.  The theory is applied to the random Green's matrix relevant to wave propagation in an ensemble of point-like scattering centers. This opens a new perspective in the study of wave diffusion, Anderson localization, and random lasing.
\end{abstract}

\maketitle

\section{Introduction}
\label{intro}

Random matrix theory is a powerful tool of modern theoretical physics \cite{mehta04}. First introduced by Wishart \cite{wishart28} and then used by Wigner to describe the statistics of energy levels in complex nuclei \cite{wigner55}, random matrices are nowadays omnipresent in physics \cite{brody81,beenakker97,guhr98,tulino04}. The majority of works --- including the seminal papers by Wigner \cite{wigner55} and Dyson \cite{dyson62} --- deal with Hermitian matrices. Hermitian matrices are, of course, of special importance in physics because of the Hermiticity of operators associated with observables in quantum mechanics. However, non-Hermitian random matrices also attracted considerable attention \cite{natano96,janik97,feinberg97}, in particular because they can be used as models for dissipative or open physical systems \cite{haake92,fyodorov97,garcia02}.

A special class of random matrices are the so-called Euclidean random matrices (ERMs) \cite{mezard99}. The elements $A_{ij}$ of a $N \times N$ Euclidean random matrix $A$ are given by a deterministic function $f$ of positions of pairs of points that are randomly distributed in a finite region $V$ of Euclidean space: $A_{ij} = f(\vec{r}_i, \vec{r}_j)$, $i = 1, \ldots, N$. Hermitian ERM models play an important role in the theoretical description of supercooled liquids \cite{mezard99,grigera03,ganter11,grigera11}, disordered superconductors \cite{chamon01}, relaxation in glasses and scalar phonon localization \cite{amir10}. They have been used as a playground to study Anderson localization \cite{bogomolny03}. A number of analytic approaches were developed to deal with Hermitian ERMs \cite{mezard99,grigera03,ganter11,grigera11,chamon01,amir10,bogomolny03,skipetrov11}.
Non-Hermitian ERMs appear in such important physical problems as Anderson localization of light \cite{rusek96} and matter waves \cite{massignan06}, random lasing \cite{pinheiro06}, propagation of light in nonlinear disordered media \cite{gremaud10}, and collective spontaneous emission of atomic systems \cite{ernst69,svid10}. However, no analytic theory is available to deal with non-Hermitian ERMs and our knowledge about their statistical properties is based exclusively on large-scale numerical simulations \cite{skipetrov11,rusek96,massignan06,pinheiro06,gremaud10}. The principal difficulties that one encounters when trying to develop a theory of non-Hermitian ERMs stem from the nontrivial statistics of their elements and the correlations between them. Both are not known analytically and are often difficult to calculate. This is in contrast with the works  \cite{haake92,fyodorov97,garcia02} where the joint probability distribution of the elements of the random matrix under study is the starting point of analysis.

In the present paper we develop an analytic theory for the density of eigenvalues of an arbitrary non-Hermitian ERM in the limit of large matrix size ($N \rightarrow \infty$). Particularly simple results are obtained for the borderline of the support of eigenvalue density on the complex plane. We illustrate the power of our approach by applying it to the `random Green's matrix' --- a matrix with elements given by the Green's function of the scalar Helmholtz equation --- that previously appeared in Refs.\ \cite{skipetrov11,rusek96,massignan06,ernst69,svid10,pinheiro06,gremaud10} but was studied only numerically up to now.
We discuss the link that exists between our calculation and the theory of wave scattering in disordered media as well as the localization properties of eigenvectors of the random Green's matrix.

\section{Foundations of the Non-Hermitian random matrix theory}
\label{nh}

The density $p(\Lambda)$ of eigenvalues $\Lambda$ of any random $N \times N$ matrix $A$ can be obtained from the resolvent
\begin{eqnarray}
g(z) = \frac{1}{N} \textrm{Tr} \left\langle \frac{1}{z-A} \right\rangle.
\label{resolvent}
\end{eqnarray}

If $A$ is Hermitian and $\Lambda$ are real, one conveniently expands $g(z)$ in series in $1/z$ in the vicinity of $|z| \rightarrow \infty$, performs the calculation using diagrammatic or any other approach, and use the result obtained after the resummation of the series at all $z$ to obtain
\begin{eqnarray}
p(\Lambda) = - \frac{1}{\pi} \lim_{\epsilon \to 0^+} \mathrm{Im}
g(\Lambda + i \epsilon).
\label{pherm}
\end{eqnarray}

For a non-Hermitian matrix $A$, however, $\Lambda$ are complex and $g(z)$ loses its analyticity inside a two-dimensional domain $\mathcal{D}$ on the complex plane where $\Lambda$ are concentrated. Thus, $g(z)$ for $z \in \mathcal{D}$ cannot be assessed by analytic continuation of its series expansion in the vicinity of $|z| \rightarrow \infty$.
A way to circumvent this problem is to double the size of the matrix and to work with a new $2N \times 2N$ matrix
\begin{equation}
A^D=
\left( \begin{array}{cc}
A & 0  \\
0& A^{\dagger}
\end{array} \right)
\label{ad}
\end{equation}
for which the generalized resolvent matrix
\begin{eqnarray}
G(Z_{\epsilon})=\frac{1}{N} \textrm{Tr}_N \left\langle \frac{1}{Z_{\epsilon} \otimes \mathbb{1}_N - A^D} \right\rangle
\label{resolvent2}
\end{eqnarray}
is safely equal to its series expansion \cite{janik97}. Here $\textrm{Tr}_N$ denotes the block trace of a $2N \times 2N$ matrix [see Eq.\ (\ref{btr}) of Appendix \ref{appendixA} for the definition] and
\begin{equation}
Z_{\epsilon}=
\left( \begin{array}{cc}
z& i \epsilon  \\
i \epsilon& z^*
\end{array} \right).
\label{ze}
\end{equation}
The resolvent $g(z)$ can be found from the diagonal elements of the $2 \times 2$ matrix obtained by taking the limit $\epsilon \to 0^+$ in Eq.\ (\ref{resolvent2}):
\begin{equation}
\lim_{\epsilon \to 0^+}G(Z_{\epsilon})=
\left[ \begin{array}{cc}
g(z) &c(z)   \\
c(z)& g(z)^*
\end{array} \right],
\label{green}
\end{equation}
and the density of eigenvalues $\Lambda$ inside its support $\mathcal{D}$ on the complex plane is \cite{janik97}
\begin{eqnarray}
p(\Lambda)= \left. \frac{1}{\pi} \frac{\partial g(z)}{\partial z^*} \right|_{z=\Lambda},
\label{pnonherm}
\end{eqnarray}
with the standard notation
$\partial/\partial z^* = \frac12 (\partial/\partial x
+ i \partial/\partial y)$ for $z = x + i y$.
The off-diagonal elements of $G$ yield the correlator of right $\ket{R_n}$ and left $\ket{L_n}$ eigenvectors of $A$ \cite{chalker98}:
\begin{eqnarray}
\mathcal{C}(z) &=&
-\frac{\pi}{N} \left\langle \sum_{n=1}^N \ps{L_n}{L_n} \ps{R_n}{R_n} \delta^{(2)}(z-\Lambda_n) \right\rangle
\nonumber \\
&=& N c(z)^2.
\label{corr}
\end{eqnarray}

\section{Non-Hermitian Euclidean random matrix theory}
\label{nherm}

All above applies to any non-Hermitian matrix $A$. Let us now make use of the fact that $A$ is an ERM with elements $A_{ij}=f(\vec{r}_i, \vec{r}_j)=\bra{ \vec{r}_i}\hat{A}\ket{ \vec{r}_j}$. Here the $N$ points $\vec{r}_i$ are randomly distributed inside some region $V$ of $d$-dimensional space with a uniform density $\rho = N/V$, and we introduced an operator $\hat{A}$ associated with the matrix $A$. A useful trick consists in changing the basis from $\{ \vec{r}_i \}$ to $\{ \psi_\alpha \}$ which is orthonormal in $V$ \cite{skipetrov11}. In a rectangular box, for example, $\ket{\psi_\alpha} = \ket{\vec{k}_\alpha}$ with
$\ps{\vec{r}}{\vec{k}_{\alpha}} = \exp(i \vec{k}_{\alpha} \vec{r})/\sqrt{V}$ \cite{skipetrov11}. For arbitrary $V$ we have
\begin{equation}
A = HTH^{\dagger},
\label{HTH}
\end{equation}
where  $H_{i\alpha}=
\ps{\vec{r}_i}{\psi_\alpha}/\sqrt{\rho}$ and $T_{\alpha\beta}=\rho\,\bra{\psi_\alpha}\hat{A}\ket{\psi_\beta}$ \footnote{In a box, $T_{\alpha\beta}$ are simply the Fourier coefficients of $f(\vec{r}_i, \vec{r}_j)$: $T_{\alpha\beta} = (\rho/V)\int_V \textrm{d}^d \vec{r}_i \int_V \textrm{d}^d \vec{r}_j f(\vec{r}_i, \vec{r}_j)
\exp[-i(\vec{k}_\alpha \vec{r}_i-
\vec{k}_\beta \vec{r}_j)]$.}.
The advantage of this representation lies in the separation of two different sources of complexity: the matrix $H$ is random but independent of the function $f$, whereas the matrix $T$ depends on $f$ but is not random.
Furthermore, if we assume that $\langle H_{i\alpha} \rangle = 0$, which in a box is obeyed for all $\alpha$ except when $\vec{k}_{\alpha} = 0$, we readily find that $H_{i\alpha}$ are identically distributed random variables with zero mean and variance equal to $1/N$.
We will assume, in addition, that $H_{i\alpha}$ are independent Gaussian random variables. This assumption largely simplifies calculations but may limit applicability of our results at high densities of points $\rho$, at least for certain types of Euclidean matrices, as we will see later.

In Appendix \ref{appendixA}, using the diagrammatic expansion of the self-energy matrix $\Sigma(Z_\epsilon) = Z_\epsilon - G(Z_\epsilon)^{-1}$, we show that, due to the representation (\ref{HTH}) and the Gaussian statistics of $H$, in the limit of large $N$, $\Sigma(Z_\epsilon)$ involves only planar rainbow-like diagrams \cite{janik97}. Summation of these diagrams yields coupled equations for operators $\hat{\Sigma}_{11}$ and $\hat{\Sigma}_{12}$ that give the elements $\Sigma_{11}=\textrm{Tr}\hat{\Sigma}_{11}/N$ and $\Sigma_{12}=\textrm{Tr}\hat{\Sigma}_{12}/N$ of the $2 \times 2$ matrix $\Sigma = \lim_{\epsilon \to 0^+}\Sigma(Z_{\epsilon})$:
\begin{eqnarray}
\hat{\Sigma}_{11}&=&(1+g\,\hat{\Sigma}_{11}+c\,\hat{\Sigma}_{12})\hat{T},
\label{self1}
\\
\hat{\Sigma}_{12}&=&(c\,\hat{\Sigma}_{11}+g^*\,\hat{\Sigma}_{12})\hat{T}^\dagger,
\label{self2}
\end{eqnarray}
where $\hat{T}=\rho\hat{A}$.
After some algebra, these equations lead to two self-consistent equations for the resolvent $g(z)$ and the eigenvector correlator $c(z)$:
\begin{eqnarray}
\frac{g^*}{|g|^2-c^2}&=&z-
\frac{1}{N}\textrm{Tr} \frac{(1-
g^*\hat{T}^{\dagger})\hat{T}}{(1-
g^*\hat{T}^{\dagger})(1- g\hat{T}) - c^2\hat{T}^{\dagger}\hat{T}},\;\;\;\;\;\;
\label{sc1}
\\
\frac{1}{|g|^2-c^2}&=& \frac{1}{N}\textrm{Tr}
\frac{\hat{T}^{\dagger}\hat{T}}{(1-
g^*\hat{T}^{\dagger})(1- g\hat{T}) - c^2\hat{T}^{\dagger}\hat{T}}.
\label{sc2}
\end{eqnarray}
%where $\hat{T}=\rho\hat{A}$.
Because $c(z)$ should vanish on the boundary $\delta\mathcal{D}$ of the support of the eigenvalue density $\mathcal{D}$, equations for $z\in\delta\mathcal{D}$ follow:
\begin{eqnarray}
&&z = \frac{1}{g}+\frac{1}{N} \textrm{Tr} \hat{S},
\label{contoura}
\\
&&\frac{1}{|g|^2} = \frac{1}{N} \textrm{Tr} \hat{S}\hat{S}^{\dagger},
\label{contourb}
\end{eqnarray}
where $\hat{S}=\hat{T}/(1-g\,\hat{T})$.

Equations (\ref{sc1}), (\ref{sc2}), (\ref{contoura}) and (\ref{contourb}) are our main results. An equation for the borderline of the support of the eigenvalue density of a non-Hermitian ERM $A$ on the complex plane $z = \Lambda$ follows from Eqs.\ (\ref{contoura}) and (\ref{contourb}) upon elimination of $g$. The density of eigenvalues $\Lambda$ inside its support $\mathcal{D}$ can be found by solving Eqs.\ (\ref{sc1}) and (\ref{sc2}) with respect to $g(z)$ and then applying Eq.\ (\ref{pnonherm}). Our analysis includes the result for Hermitian ERMs as a special case: if $A$ is Hermitian, then $\Sigma$ is diagonal and the support of the eigenvalue density shrinks to a segment on the real axis. Equation (\ref{contoura}) then allows one to solve for $g(z)$. This result for Hermitian matrices coincides with the one found in Ref.\ \cite{skipetrov11} using a different approach.

The solution of Eqs.\ (\ref{sc1}), (\ref{sc2}), (\ref{contoura}) and (\ref{contourb}) for a given matrix $A$ is greatly facilitated by a suitable choice of the basis in which traces appearing in these equations are expressed. In addition to $\{ \vec{r} \}$ and $\{ \vec{k}_{\alpha} \}$, a bi-orthogonal basis of right $\ket{\mathcal{R}_{\alpha}}$ and left $\ket{\mathcal{L}_{\alpha}}$ eigenvectors of $\hat{T}$ can be quite convenient. The right eigenvector $\ket{\mathcal{R}_{\alpha}}$ obeys
\begin{eqnarray}
\bra{\vec{r}} \hat{T} \ket{\mathcal{R}_{\alpha}} =
\rho \int_V\textrm{d}^d \vec{r'}f(\vec{r}, \vec{r'}) \mathcal{R}_\alpha(\vec{r'}) = \mu_\alpha \mathcal{R}_\alpha(\vec{r}),
\label{right}
\end{eqnarray}
where $\mu_{\alpha}$ is the eigenvalue corresponding to the eigenvector $\ket{\mathcal{R}_{\alpha}}$. The traces appearing in Eqs.\ (\ref{contoura}) and (\ref{contourb}) can be expressed as
\begin{eqnarray}
\textrm{Tr} \hat{S} &=&
\sum_{\alpha} \bra{\mathcal{L}_{\alpha}} \hat S \ket{\mathcal{R}_{\alpha}} = \sum_{\alpha} \frac{\mu_\alpha}{1-g\mu_\alpha},
\label{trs}
\\
\textrm{Tr} \hat{S}\hat{S}^{\dagger} &=&
\sum_{\alpha,\beta}
\frac{\mu_\alpha \mu_\beta^* \ps{\mathcal{L}_\alpha}{\mathcal{L}_\beta}
\ps{\mathcal{R}_\beta}
{\mathcal{R}_\alpha}}{(1-g\mu_\alpha)(1-g\mu_\beta)^*},
\label{trss}
\end{eqnarray}
respectively.

\section{Example: random Green's matrix}
\label{greenmatrix}

Let us now illustrate the power of the above general analysis on the example of the random Green's matrix
\begin{equation}
A_{ij}=(1-\delta_{ij})\frac{\exp(i k_0| \vec{r}_i-\vec{r}_j|)}{k_0|\vec{r}_i-\vec{r}_j|},
\label{ERMGreen}
\end{equation}
where $k_0 = 2 \pi/\lambda_0$ and $\lambda_0$ is the wavelength. We assume that the $N$ points $\vec{r}_i$ are chosen randomly inside a three-dimensional ($d = 3$) sphere of radius $R$. This non-Hermitian ERM is of special importance in the context of wave propagation in disordered media because its elements are proportional to the Green's function of Helmholtz equation, with $\vec{r}_i$ that may be thought of as positions of point-like scattering centers. It previously appeared in Refs.\ \cite{skipetrov11,rusek96,massignan06,ernst69,svid10,pinheiro06,gremaud10}, but was studied only by extensive numerical simulations, except in Ref.\ \cite{svid10} where analytic results were obtained in the infinite density limit.

For each realization of the random matrix (\ref{ERMGreen}), its eigenvalues $\Lambda_n$ obey \cite{skipetrov11}
\begin{eqnarray}
\sum\limits_{n=1}^N \Lambda_n = 0,
\;\;\; \textrm{Im} \Lambda_n > -1.
\label{conditions}
\end{eqnarray}
Very generally, the eigenvalue density of the matrix defined by Eq.\ (\ref{ERMGreen}) depends on two dimensionless parameters: the number of points per wavelength cubed $\rho \lambda_0^3$ and the second moment of $|\Lambda|$ calculated in the limit of low density: $\langle |\Lambda|^2 \rangle = \gamma = 9N/8(k_0 R)^2$. Even though the latter result for $\langle |\Lambda|^2 \rangle$ can be rigourously justified only in the limit of low density $\rho \lambda_0^3 \ll 1$, it holds approximately up to densities as high as $\rho \lambda_0^3 \sim 100$. We will see from the following that the two parameters $\rho \lambda_0^3$ and $\gamma$ control different properties of the eigenvalue density.

\subsection{Borderline of the eigenvalue domain}

%%%%%%%%%%%%%  FIG1 %%%%%%%%%
\begin{figure*}[!t]
\centering{
\includegraphics[angle=0,width=0.9\textwidth]{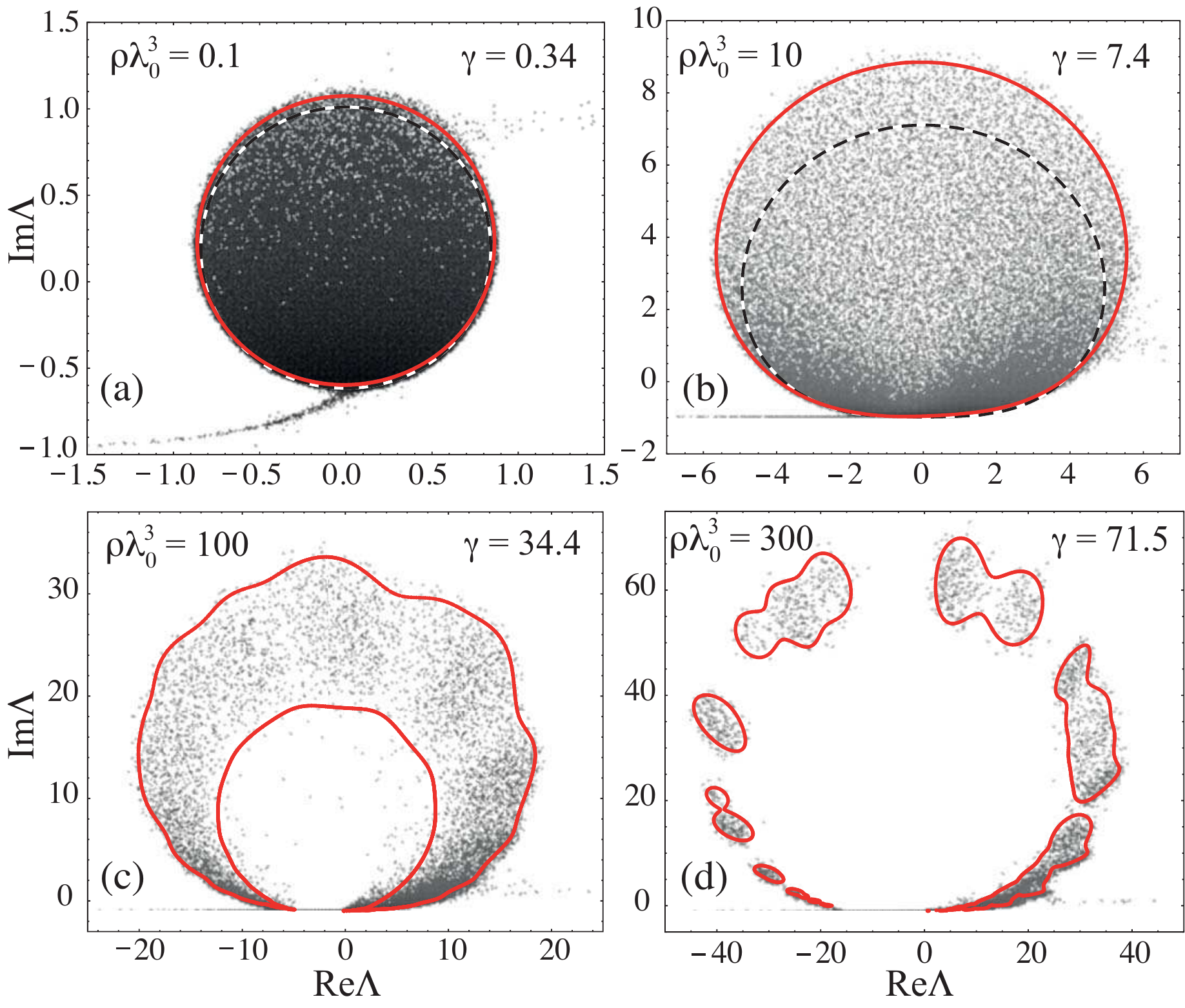}
\caption{
\label{fig1}
Density plots of the logarithm of eigenvalue density of the $N \times N$ random Green's matrix (\ref{ERMGreen}) obtained by numerical diagonalization of 10 realizations of the matrix for $N = 10^4$. The solid lines represent the borderlines of the support of eigenvalue density following from Eq.\ (\ref{contour2}) in panels (a) and (b) and from Eqs.\ (\ref{contourT3}) and (\ref{contourT4}) of Appendix \ref{appendixC} in panels (c) and (d). The dashed lines show the diffusion approximation (\ref{diffusion}).}}
\end{figure*}
%%%%%%%%%%%%%%%%%%%%%%%%%%

We first focus on the borderline of the support of eigenvalues which is easier to visualize.
In Fig.\ \ref{fig1} we present a comparison of the solutions of Eqs.\ (\ref{contoura}) and (\ref{contourb}) with results of numerical diagonalization of the matrix (\ref{ERMGreen}) for $k_0 R \gg 1$. At low density $\rho\lambda_0^3 \lesssim 10$, a sufficiently accurate solution of Eqs. (\ref{contoura}) and (\ref{contourb}) can be obtained in the $\ket{\vec{r}}$ representation, in which
\begin{eqnarray}
\bra{ \vec{r}}\hat{S}\ket{ \vec{r'}} \simeq \rho \frac{\exp(i \kappa | \vec{r}-\vec{r'}|)}{k_0 | \vec{r}-\vec{r'}|},
\label{lowrhos}
\end{eqnarray}
with $\kappa(g)=k_0\sqrt{1+g\rho\lambda_0^3/2\pi^2}$. In Appendix \ref{appendixB} we show that this leads to a borderline equation
\begin{equation}
|\Lambda|^2 = 2 \gamma h \left[
2\textrm{Im} \kappa\left( 1/\Lambda \right) R
\right],
\label{contour2}
\end{equation}
where
\begin{eqnarray}
h(x) = \frac{3-6x^2+8x^3-3(1+2x)e^{-2x}}{6x^4}.
\label{h}
\end{eqnarray}
For  $\rho \lambda_0^3 \lesssim 10$, a simpler equation
\begin{eqnarray}
|\Lambda|^2 \simeq 2 \gamma h \left( -8\gamma \frac{\textrm{Im}\Lambda}{3|\Lambda|^2} \right)
\label{contour3}
\end{eqnarray}
yields satisfactory results as well. For $\gamma \ll 1$, the density of eigenvalues is roughly uniform within a circular domain of radius $\sqrt{2\gamma}$, see Fig.\ \ref{fig1}(a). The domain grows in size and shifts up upon increasing $\gamma$. At $\gamma \gtrsim 1$ it starts to `feel' the `wall' $\textrm{Im} \Lambda = -1$ and deforms [Fig.\ \ref{fig1}(b)].

The approximate equation (\ref{contour2}) for the borderline of the support of eigenvalue density yields a closed line on the complex plane until $\rho\lambda_0^3 \simeq 30$, after which the line opens from below. This signals that an important change in behavior might be expected at this density. And indeed, we observe that a `hole' opens in the eigenvalue density for $\rho\lambda_0^3 \gtrsim 30$. As we see in Fig.\ \ref{fig1}(c), this hole is perfectly described by our Eqs.\ (\ref{contoura}) and (\ref{contourb}) which we now solve in the basis of eigenvectors of the operator $\hat{T}$. Eigenvalues and eigenvectors of $\hat{T}$ can be found analytically \cite{svid10}. As we discuss in Appendix \ref{appendixC}, this allows for an exact solution of Eqs.\ (\ref{contoura}) and (\ref{contourb}). Finally, at very high density the crown formed by the eigenvalues blows up in spots centered around the eigenvalues $\mu_\alpha$ of $\hat{T}$, as we show in Fig.\ \ref{fig1}(d). When the density is further increased, the eigenvalues $\Lambda_n$ of $A$ become equal to the eigenvalues $\mu_{\alpha}$ of $\hat{T}$. They then fall on the circular line given by Eq.\ (\ref{largerho}) and the problem looses its statistical nature. As follows from our analysis, the parameter $\gamma$ controls the overall extent of the support of eigenvalue density $\mathcal{D}$ on the complex plane, whereas its structure depends also on the density $\rho\lambda_0^3$. At fixed $\gamma$, $\mathcal{D}$ goes through a transition from a disk-like to an annulus-like shape, and eventually splits into multiple disconnected spots upon increasing $\rho\lambda_0^3$.
The transition from disk-like to the annulus-like shape is reminiscent of the disk-annulus transition in the eigenvalue distribution of rotationally invariant non-Hermitian random matrix ensembles \cite{feinberg97}.

An important additional feature of the numerical results in Fig.\ \ref{fig1} that is not described by our Eqs.\ (\ref{contoura}) and (\ref{contourb}) is the eigenvalues that concentrate around the two hyperbolic spirals, $|\Lambda|=1/\arg{\Lambda}$ and its reflection through the origin. These spirals correspond to the two eigenvalues $\pm A_{12}$ of the matrix (\ref{ERMGreen}) for $N = 2$ \cite{rusek96,skipetrov11}. The eigenvectors corresponding to these eigenvalues are localized on pairs of very close points. From numerical results for $N \leq 10^4$, we estimate their statistical weight to be important at large densities, of the order of $1 - \mathrm{const}/(\rho \lambda_0^3)^p$ with $p \sim 1$. This is consistent with the estimation of the number of subradiant states in a large atomic cloud by Ernst \cite{ernst69}. At large densities, the absolute majority of the lacking eigenvalues fall very close to the axis $\mathrm{Im} \Lambda = -1$, in the `gap' that opens in the eigenvalue distribution following from our theory on the left from $\mathrm{Re} \Lambda = 0$ [see Figs.\ \ref{fig1}(c) and (d)]. The lack of the spiral branches of $p(\Lambda)$ in our theory can be traced back to the assumption of statistical independence of elements of the matrix $H$ in Eq.\ (\ref{HTH}). It does not affect the excellent agreement of the borderline of the rest of the eigenvalue domain with numerical results.

\subsection{Mapping to the scattering theory}

We now want to introduce an interesting mapping between our results for the random Green's matrix (\ref{ERMGreen}) and the problem of multiple scattering of waves by $N$ resonant point-like scatterers. The latter problem is described by the Helmholtz equation associated with a fictitious Hamiltonian
\begin{eqnarray}
\hat{H} = -\nabla^2 +v(k_0)\sum_{i=1}^N
\delta^{(3)}( \hat{\vec{r}}- \vec{r}_i).
\label{fictious}
\end{eqnarray}
The retarded free-space Green's function corresponding to $v(k_0) = 0$,
\begin{eqnarray}
\hat{\mathcal{G}}_0 = \frac{1}{k_0^2 + i\epsilon + \nabla^2},
\label{retarded}
\end{eqnarray}
is simply proportional to the matrix (\ref{ERMGreen}):
\begin{eqnarray}
(\mathcal{G}_0)_{ij}=\bra{ \vec{r}_i}\hat{\mathcal{G}}_0\ket{ \vec{r}_j}=-\frac{k_0}{4\pi} A_{ij}.
\label{g0ij}
\end{eqnarray}
Expanding the Green's function
\begin{eqnarray}
\hat{\mathcal{G}} = \frac{1}{k_0^2 + i\epsilon - \hat{H}}
\label{greenh}
\end{eqnarray}
in Born series, we get
\begin{eqnarray}
\mathcal{G} = \frac{1}{\mathcal{G}^{-1}_0-t},
\label{greenh2}
\end{eqnarray}
where $t$ is the scattering matrix of an individual scatterer defined by \cite{sheng06}
\begin{align}
t \delta^{(3)}(\hat{\vec{r}} - \vec{r}_i) =
\left[
 v(k_0) + v(k_0)\delta^{(3)}(\hat{\vec{r}} - \vec{r}_i) \hat{{\cal G}}_0 t 
 \right]
 \delta^{(3)}(\hat{\vec{r}} - \vec{r}_i).
\label{tmatrix}
\end{align}

At $\vec{r}_i$, the intensity of a wave emitted by a point source located at $\vec{r}_j$ is $I_{ij} = |\mathcal{G}_{ij}|^2$, where
$\mathcal{G}_{ij} = \bra{\vec{r}_i} \hat{\mathcal{G}} \ket{\vec{r}_j}$. Let us introduce $I(t) = \sum_{i \neq  j} I_{ij}$, where we emphasize that $I$ depends on $t$. It can be readily written as
\begin{eqnarray}
I(t) = \textrm{Tr} \frac{1}{(t-\mathcal{G}_0^{-1})(t-\mathcal{G}_0^{-1})^{\dagger}}.
\label{it}
\end{eqnarray}
This is to be compared with the expression for the correlator of right and left eigenvectors of an arbitrary matrix $A$ following from Eq.\ (\ref{green}):
\begin{eqnarray}
c(z) = -\lim_{\epsilon \to 0^+} \frac{i\epsilon}{N}
\textrm{Tr} \left\langle \frac{1}{(z-A)(z-A)^{\dagger}+\epsilon^2} \right\rangle.
\label{corr2}
\end{eqnarray}
For $A = \mathcal{G}_0^{-1}$ and $z = t$ we thus have
\begin{eqnarray}
c(t) = -\lim_{\epsilon \to 0^+} \frac{i\epsilon}{N}
\langle I(t) \rangle.
\label{ci}
\end{eqnarray}
This should become different from zero when $t$ enters the support of the eigenvalue density of $\mathcal{G}_0^{-1}$ or, equivalently, when $1/t$ enters the support of the eigenvalue density of $\mathcal{G}_0$. The only way to obtain $c(t) \ne 0$ for $\epsilon \to 0^+$ is to make $I(t)$ diverge. In the framework of our linear model of scattering, this can be achieved by realizing a random laser \cite{wiersma08}. We thus come to the surprising conclusion that finding the borderline of the support of the eigenvalue density $p(\Lambda)$ of the $N \times N$ Green's matrix (\ref{ERMGreen}) is mathematically equivalent to calculating the threshold for random lasing in an ensemble of $N$ identical point-like scatterers with scattering matrix $t = -4\pi/k_0 \Lambda$. In the diffusion approximation, for example, the threshold of such a random laser can be found as in Ref.\ \cite{perez09}. This leads to the following equation for the borderline \footnote{We use the extrapolation length $z_0 = 2/3$ \cite{sheng06} instead of $z_0 = 0.71$ in Ref.\ \cite{perez09}}:
\begin{equation}
|\Lambda|^2 = \frac{8\gamma}{\sqrt{3}\pi} \sqrt{1+\textrm{Im} \Lambda}
\left(1 + \frac{|\Lambda|^2}{|\Lambda|^2+4 \gamma} \right).
\label{diffusion}
\end{equation}
We show this equation in Figs.\ \ref{fig1}(a) and (b) by dashed lines. As expected, it gives satisfactory results only in the weak scattering regime $\rho \lambda_0^3 \lesssim 10$ and at large optical thickness $b = 2 R/\ell = 16 \gamma/3|\Lambda|^2 \gg 1$, where $\ell = 4\pi/\rho |t|^2$ is the mean free path. In contrast, our Eqs.\ (\ref{contoura}) and (\ref{contourb}) apply at any $\rho \lambda_0^3$ and $b$. These equations can therefore serve as a benchmark for theories of multiple scattering.

\subsection{Eigenvalue density}

%%%%%%%%%%%%%  FIG3 %%%%%%%%%%
\begin{figure}[!t]
\centering{
\includegraphics[angle=0,width=\columnwidth]{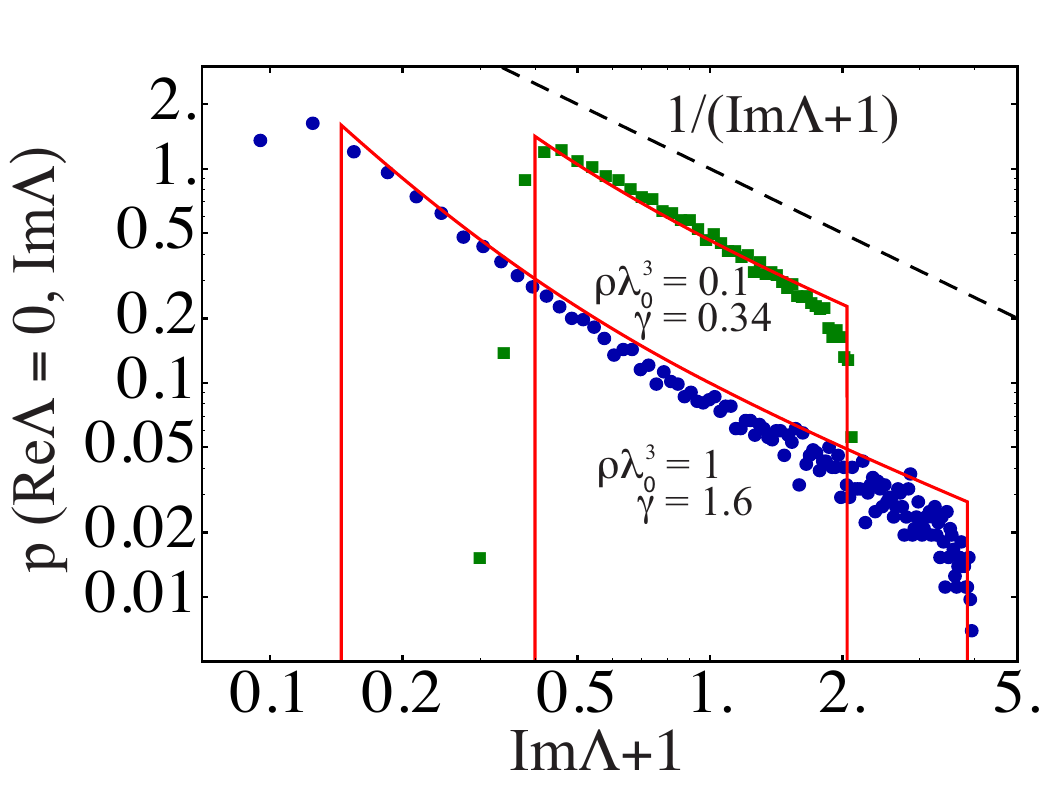}
\caption{
\label{fig3}
Cuts of the eigenvalue density $p(\Lambda)$ of the Green's matrix (\ref{ERMGreen}) along the imaginary axis $\mathrm{Re} \Lambda = 0$.  Numerical simulations (symbols) are compared with our analytical solution (\ref{resolvent}).}}
\end{figure}
%%%%%%%%%%%%%%%%%%%%%%%%%%

Let us now analyze the shape of the eigenvalue density $p(\Lambda)$ inside its support $\mathcal{D}$ using Eqs.\ (\ref{sc1}) and (\ref{sc2}). Very generally, $p(\Lambda)$ is roughly symmetric with respect to the line $\mathrm{Re} \Lambda = 0$ and decays with $\mathrm{Im} \Lambda$. A particular feature of $p(\Lambda)$ that was studied previously is the behavior of the marginal probability density of $\mathrm{Im} \Lambda$. Pinheiro {\em et al.} \cite{rusek96} observed $p(\textrm{Im} \Lambda) \propto 1/(\textrm{Im} \Lambda+1)$ in numerical simulations at high density and conjectured that it was a signature of Anderson localization of waves in the corresponding point-scatterer model. To test this conjecture, we analyze $p(\Lambda)$ at low densities $\rho \lambda_0^3 \lesssim 1$, for which no Anderson localization is expected. An  approximate solution of Eqs.\ (\ref{sc1}) and (\ref{sc2}) in this regime can be obtained by neglecting the term $c^2 \hat{T}^{\dagger} \hat{T}$ in their denominators:
\begin{eqnarray}
g(z) = \frac{z^* - \frac{1}{N} \mathrm{Tr} \hat{S}^{\dagger}}{\frac{1}{N} \mathrm{Tr} \hat{S} \hat{S}^{\dagger}}.
\label{resolvent}
\end{eqnarray}
Traces in this equation can be explicitly calculated using Eq.\ (\ref{lowrhos}) valid at low densities, as we show in Appendix \ref{appendixB}. The eigenvalue density $p(\Lambda)$ is then found by applying Eq.\ (\ref{pnonherm}). In Fig.\ \ref{fig3} we show cuts of $p(\Lambda)$ along the imaginary axis $\mathrm{Re} \Lambda = 0$. We clearly observe that $p(\mathrm{Re} \Lambda = 0, \mathrm{Im} \Lambda)$ decays as  $1/(\textrm{Im} \Lambda+1)$, even though the density of points $\rho \lambda_0^3$ is too low to bring the system to the Anderson localization transition. However, the power-law decay becomes clearly visible in the marginal distribution $p(\textrm{Im} \Lambda)$ only when the support of $p(\textrm{Im} \Lambda)$ is sufficiently wide, i.e. for $\gamma \gtrsim 1$. Otherwise, it is `spoiled' by the circular shape of the support of $p(\Lambda)$ and $p(\textrm{Im} \Lambda)$ follows the Marchenko-Pastur law \cite{skipetrov11}. Because the condition $\gamma \gtrsim 1$ can be obeyed at any, even very low density by just increasing the number of points $N$, it seems that no direct link can be established between the power-law decay of $p(\textrm{Im} \Lambda)$ and Anderson localization.

\subsection{Anderson localization}

\begin{figure}[!t]
\centering{
\includegraphics[angle=0,width=\columnwidth]{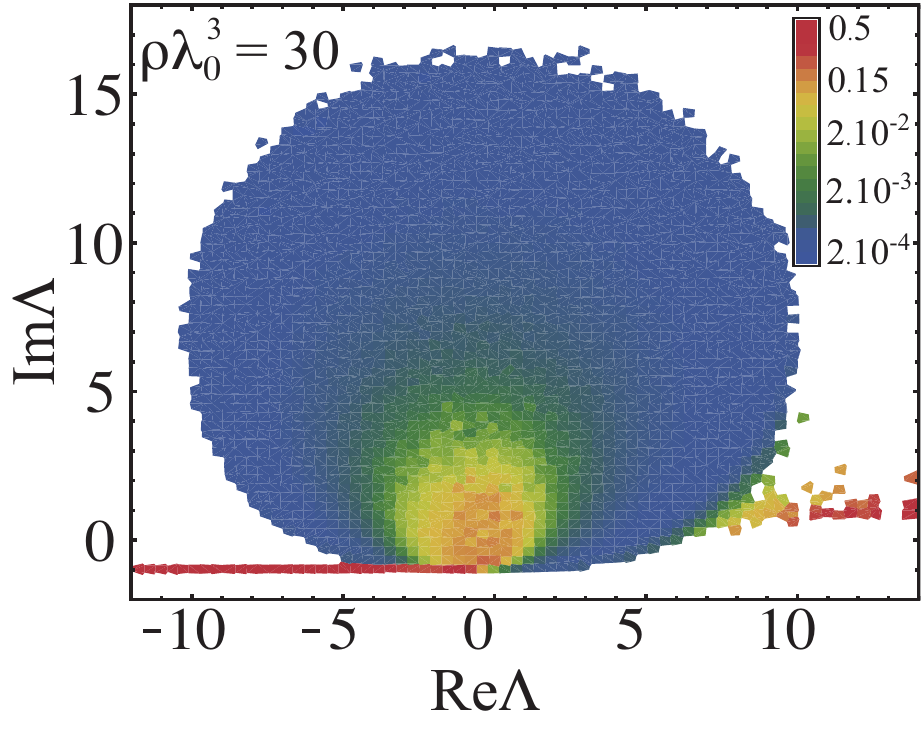}
\caption{
\label{fig4}
Density plot of the logarithm of the average inverse participation ratio of eigenvectors of the Green's matrix (\ref{ERMGreen}). To obtain this plot, we found eigenvalues of 10 different random realizations of $10^4 \times 10^4$ Green's matrix numerically, computed their IPRs using Eq.\ (\ref{ipr}), and then determined $\mathrm{IPR}(\Lambda)$ by integrating Eq.\ (\ref{ipr2}) over a small area $(\Delta \Lambda)^2$ around $\Lambda$, for a grid of $\Lambda$'s on the complex plane.}}
\end{figure}

It should be stressed here that Anderson localization --- the localization of eigenvectors in space due to disorder --- is a property of \textit{eigenvectors} $\ket{R_n}$ of the matrix (\ref{ERMGreen}), whereas our study in this paper concerns its \textit{eigenvalues} $\Lambda_{n}$. It is not clear \textit{a priori} if any sign of Anderson localization should (and could) be visible in the density of eigenvalues $p(\Lambda)$. To elaborate on this issue, we analyze the eigenvectors of the matrix (\ref{ERMGreen}). To determine if an eigenvector $\ket{R_n}$ is localized or not, we compute its inverse participation ratio (IPR):
\begin{equation}
\textrm{IPR}_{n} = \frac{\sum_{i=1}^N |R_{n}(\vec{r}_i)|^4}{\left[\sum_{i=1}^N |R_{n}(\vec{r}_i)|^2 \right]^2}.
\label{ipr}
\end{equation}
An eigenvector extended over all $N$ points is characterized by $\mathrm{IPR} \sim 1/N$, whereas an eigenvector localized on a single point has $\mathrm{IPR} = 1$. The average value of IPR corresponding to eigenvectors with eigenvalues in the vicinity of $\Lambda$ can be defined as
\begin{equation}
\textrm{IPR}(\Lambda) = \frac{1}{p(\Lambda)}
\left\langle
\sum\limits_{n=1}^{N} \mathrm{IPR}_{n}\;
\delta^2(\Lambda - \Lambda_{n})
\right \rangle,
\label{ipr2}
\end{equation}
where averaging is over all possible configurations of $N$ points in a sphere. Our numerical analysis of the average IPR defined by this equation reveals the following scenario. At low density $\rho \lambda_0^3 \lesssim 10$, $\mathrm{IPR} \simeq 2/N$ for all eigenvectors except those corresponding to the eigenvalues that belong to spiral branches in Fig.\ \ref{fig1}(a) and (b) for which $\mathrm{IPR} \simeq \frac12$. These states are localized on pairs of points that are very close together and correspond to proximity resonances \cite{rusek96} that do not require a large optical thickness to build up. The prefactor $2$ in the result for $\mathrm{IPR}$ of extended eigenvectors is due to the Gaussian statistics of eigenvectors at low densities. For $\rho \lambda_0^3 \gtrsim 10$, IPR starts to grow in a roughly circular domain in the vicinity of $\Lambda = 0$ and reaches maximum values $\sim 0.1$ at $\rho \lambda_0^3 \simeq 30$ [see Fig.\ \ref{fig4}]. Contrary to common belief \cite{rusek96}, neither localized states necessarily have $\mathrm{Im} \Lambda$ close to $-1$, nor states with $\mathrm{Im} \Lambda \simeq -1$ are always localized, as can be seen from Fig.\ \ref{fig4}. For $\rho \lambda_0^3 > 30$, the localized states start to disappear and a hole opens in the eigenvalue density. It is quite remarkable that the opening of the hole in $p(\Lambda)$ proceeds by disappearance of localized states (i.e., of states with $\mathrm{IPR} \gg 1/N$).

%%%%%%%%%%%%%  FIG4 %%%%%%%%%%
%%%%%%%%%%%%%%%%%%%%%%%%%%

\section{Conclusion}
\label{concl}

We derived equations for the resolvent $g(z)$ and the correlator $c(z)$ of right and left eigenvectors of an arbitrary $N \times N$ non-Hermitian Euclidean random matrix in the limit of $N \to \infty$. These equations allow us to analyze the borderline of the support of eigenvalues $\Lambda$ by looking for a contour on the complex plane on which $c(z) = 0$, as well as the full probability density $p(\Lambda)$ inside this contour by solving for $g(z)$. To give an example of application of our general results to a particular physical problem, we studied the eigenvalue density of the random Green's matrix (\ref{ERMGreen}). An entry $A_{ij}$ of this matrix is equal to the Green's function of the scalar Helmholtz equation between two points $\vec{r}_i$ and $\vec{r}_j$ chosen among $N$ points randomly distributed in a sphere. We showed that finding the borderline of the support of the eigenvalue density of the Green's matrix is mathematically equivalent to calculating the threshold for random lasing in an ensemble of $N$ identical point-like scatterers. Finally, we discussed manifestations of Anderson localization in the properties of this matrix and challenged the link that was previously proposed between Anderson localization and the power-law decay of the marginal probability density.
%$p(\mathrm{Im} \Lambda) \propto 1/(\mathrm{Im} \Lambda + 1)$.

\acknowledgements
This work was supported by the French ANR (project no. 06-BLAN-0096 CAROL).

\appendix
%%%%%%%%%%%%%%   APPENDIX A   %%%%%%%%%%%%%%
\section{Derivation of self-consistent equations for the resolvent and the eigenvector correlator }
\label{appendixA}

The purpose of this Appendix is to derive Eqs. (\ref{sc1}) and (\ref{sc2}) of the main text. We start by expanding the $2\times 2$ resolvent matrix  $G(Z_{\epsilon})$ defined by Eq.\ (\ref{resolvent2}) in series in $1/\mathcal{Z}_{\epsilon} = (1/Z_{\epsilon}) \otimes \mathbb{1}_N $:
\begin{align}
G(Z_{\epsilon})
&=\left( \begin{array}{cc}
G_{11}^\epsilon & G_{12}^\epsilon  \\
G_{12}^\epsilon & G_{11}^{\epsilon *}
\end{array} \right)&
\nonumber
\\
&=\frac{1}{N} \textrm{Tr}_{N}\left\langle
\frac{1}{\mathcal{Z}_{\epsilon}}+
\frac{1}{\mathcal{Z}_{\epsilon}}\,A^D\,
\frac{1}{\mathcal{Z}_{\epsilon}}+\ldots\right\rangle,
\label{Gseries}
\end{align}
where the averaging $\left< \ldots \right>$ is performed over the ensemble of matrices $H$ entering the representation (\ref{HTH}) of the matrix $A$. The block trace $\mathrm{Tr}_N X$ of an arbitrary $2N \times 2N$ matrix $X$ is defined by separating $X$ in four $N \times N$ blocks $X_{11}$, $X_{12}$, $X_{21}$, $X_{22}$ and taking the trace of each of the latter separately:
\begin{eqnarray}
\nonumber
\mathrm{Tr}_N X &=& \mathrm{Tr}_N \left( \begin{array}{cc}
X_{11} & X_{12}  \\
X_{21} & X_{22}
\end{array} \right)
\\
&=& \left( \begin{array}{cc}
\mathrm{Tr} X_{11} & \mathrm{Tr} X_{12}  \\
\mathrm{Tr} X_{21} & \mathrm{Tr} X_{22}
\end{array} \right).
\label{btr}
\end{eqnarray}

As explained in the main text, we assume that $H$ has independent identically distributed complex entries that obey circular Gaussian distribution. Using the properties of Gaussian random variables, the result of averaging in Eq.\ (\ref{Gseries}) can be expressed through pairwise contractions
\begin{equation}
 \left\langle H_{i\alpha}H^\dagger_{\beta j}\right\rangle = \frac{1}{N}\delta_{ij}\delta_{\alpha \beta}=\left\langle H^\dagger_{\alpha i}H_{j \beta}\right\rangle.
 \label{Contraction}
\end{equation}

\begin{figure}[h!]
\centering{
\includegraphics[angle=0,width=0.7\columnwidth]{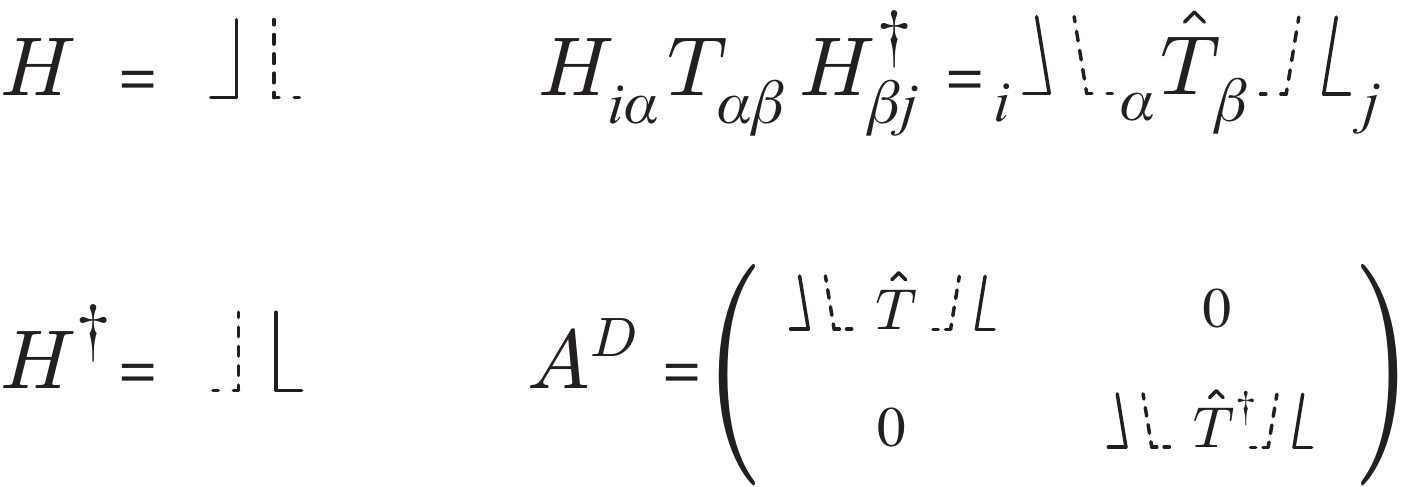}
\caption{
\label{figA1}
Diagrammatic representations of the matrices $H$, $H^{\dagger}$, $A=HTH^{\dagger}$ and $A^D$. Full and dashed lines propagate in the bases $\{\mathbf{r}_i\}$ and $\{\psi_\alpha\}$, respectively.
}}
\end{figure}

To evaluate efficiently the weight of different terms that arise in the calculation, it is convenient to introduce diagrammatic notations. First, the matrices $H$, $H^{\dagger}$, $A$ and $A^D$ will be represented as shown in Fig. \ref{figA1}.

%%%%%FIGA1%%%%%
%%%%%%%%%%%%%

The `propagator' $1/\mathcal{Z}_{\epsilon}$ will be depicted by
\begin{equation}
\frac{1}{Z_\epsilon}
=
\left( \begin{array}{cc}
\frac{1}{z} & -\frac{i\epsilon}{\vert z\vert^2}   \\
 -\frac{i\epsilon}{\vert z\vert^2}&\frac{1}{z^*}
\end{array} \right)
=
\left( \begin{array}{cc}
\overline{\mbox{\scriptsize 1\;\;\;\;1}} &\overline{\mbox{\scriptsize 1\;\;\;\;2}}   \\
\overline{\mbox{\scriptsize 2\;\;\;\;1}}&\overline{\mbox{\scriptsize 2\;\;\;\;2}}
\end{array} \right).
\label{InvZ}
\end{equation}

\begin{figure}[h!]
\vspace{2mm}
\centering{
\includegraphics[angle=0,width=0.5\columnwidth]{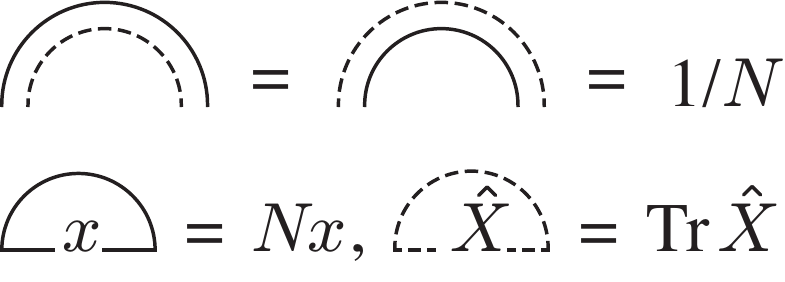}
\caption{
\label{figA2}
Diagrammatic notation for pairwise contractions (\ref{Contraction}) and loop diagrams for any scalar $x$ in the basis $\{\mathbf{r}_i\}$, and for any operator $\hat{X}$ in an arbitrary basis $\{\psi_\alpha\}$.}}
\end{figure}

Each contraction (\ref{Contraction}) brings a factor $1/N$, and each loop corresponding to taking the trace of a matrix brings a factor $N$, see Fig. \ref{figA2}.

%%%%%FIGA2%%%%%
%%%%%%%%%%%%%

\begin{figure}[h]
\centering{
\includegraphics[angle=0,width=0.9\columnwidth]{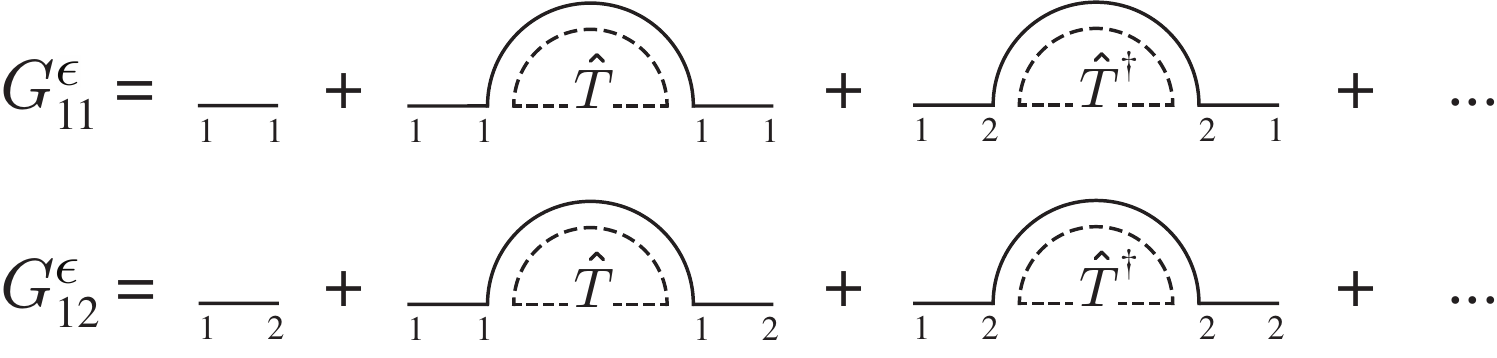}
\caption{
\label{figS1}
Diagrammatic expansion of the two independent elements of the matrix $G(Z_\epsilon)$. }}
\end{figure}

In the limit $N \rightarrow \infty$, only the diagrams that contain as many loops as contractions will survive. These diagrams are those where full and dashed lines do not cross.
Therefore, the leading order expansion of the resolvent (\ref{Gseries}) involves only diagrams which are planar and look like rainbows. Such diagrams appear, for example, in Fig.\ \ref{figS1}, where we show the beginning of the expansion of the two independent elements of $G(Z_\epsilon)$.

%%%%%FIGS1%%%%%
%%%%%%%%%%%%%

\begin{figure*}[t]
\centering{
\includegraphics[angle=0,width=0.9\textwidth]{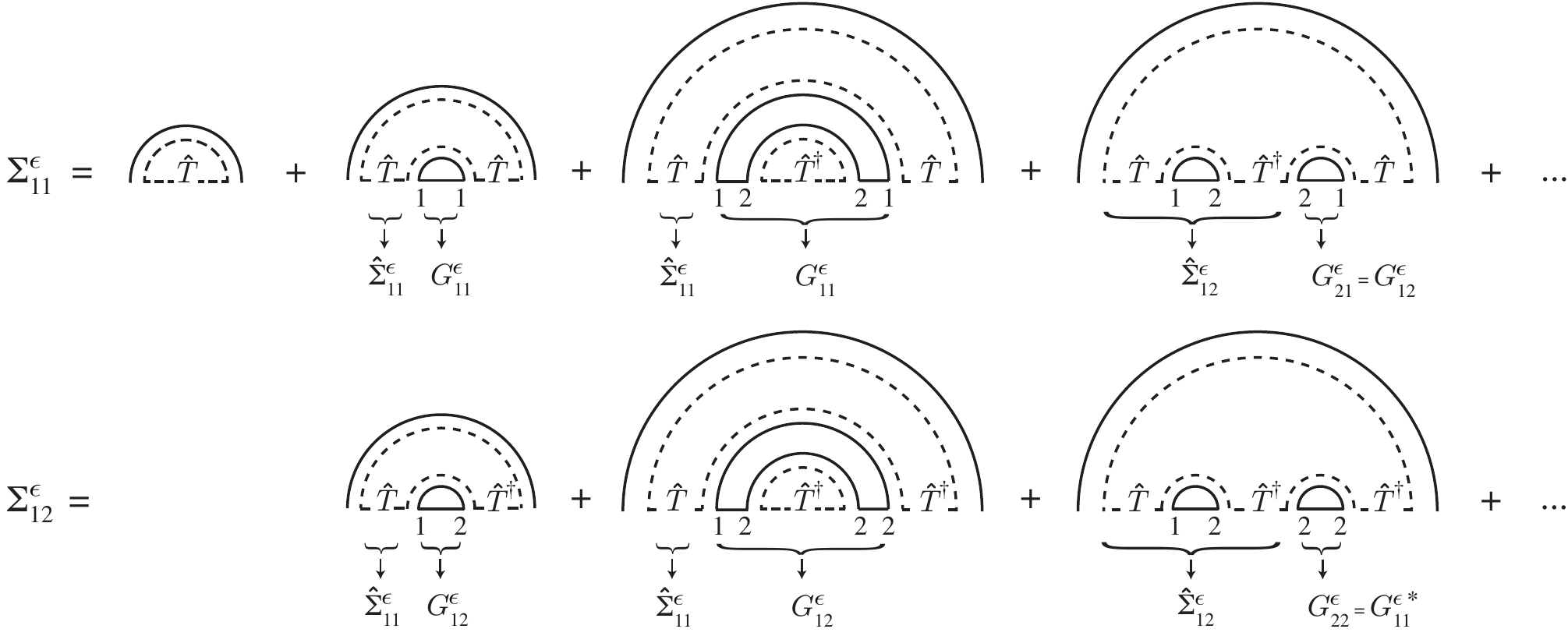}
\caption{
\label{figS2}
Diagrammatic expansion of the two independent elements of the self-energy $\Sigma(Z_\epsilon)$. Braces with arrows denote parts of diagrams that are beginning of diagrammatic expansions of the quantities which the arrows point to.}}
\end{figure*}

\begin{figure}[h!]
\centering{
\includegraphics[angle=0,width=0.6\columnwidth]{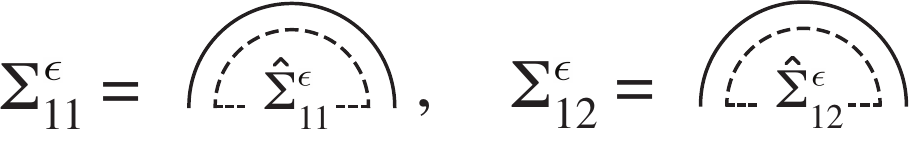}
\caption{
\label{figA3}
The elements $\Sigma_{11}^\epsilon$ and $ \Sigma_{12}^\epsilon$  of the matrix $\Sigma(Z_{\epsilon})$ can be written as traces of operators $\hat{\Sigma}_{11}^\epsilon$ and $\hat{\Sigma}_{12}^\epsilon$ that appear in Fig.\ \ref{figS2}: $\Sigma_{11}^\epsilon=\textrm{Tr}\hat{\Sigma}_{11}^\epsilon/N$ and $\Sigma_{12}^\epsilon=\textrm{Tr}\hat{\Sigma}_{12}^\epsilon/N$. }}
\end{figure}

In the standard way, rather than summing up the diagrams for the resolvent, we introduce the $2\times 2$ self-energy matrix
\begin{equation}
\Sigma(Z_\epsilon) = Z_\epsilon - G(Z_\epsilon)^{-1}=\left( \begin{array}{cc}
\Sigma_{11}^\epsilon & \Sigma_{12}^\epsilon  \\
\Sigma_{12}^\epsilon & \Sigma_{11}^{\epsilon *}
\end{array} \right).
 \end{equation}
It is equal to the sum of all one-particle irreducible diagrams contained in
\begin{equation}
Z_\epsilon G(Z_\epsilon)Z_\epsilon=\frac{1}{N} \textrm{Tr}_N \left\langle A^D+A^D\frac{1}{\mathcal{Z}_{\epsilon}}A^D+\ldots\right\rangle.
\label{Selfseries}
\end{equation}
The first dominant terms that appear in the expansion of the two matrix elements $\Sigma_{11}^\epsilon$ and $\Sigma_{12}^\epsilon$ are represented in Fig. \ref{figS2}.
%%%%%FIGS2%%%%%
%%\begin{widetext}
%%\end{widetext}
%%%%%%%%%%%%%
In the two series of Fig. \ref{figS2} we recognize, under a pairwise contraction, the matrix elements $G_{11}^\epsilon$ and $G_{12}^\epsilon$ depicted in Fig. \ref{figS1}, as well as the two operators $\hat{\Sigma}_{11}^\epsilon$ and  $\hat{\Sigma}_{12}^\epsilon$ defined in Fig. \ref{figA3}.

%%%%%FIGA3%%%%%
%%%%%%%%%%%%%

\begin{figure}[h!]
\centering{
\includegraphics[angle=0,width=0.9\columnwidth]{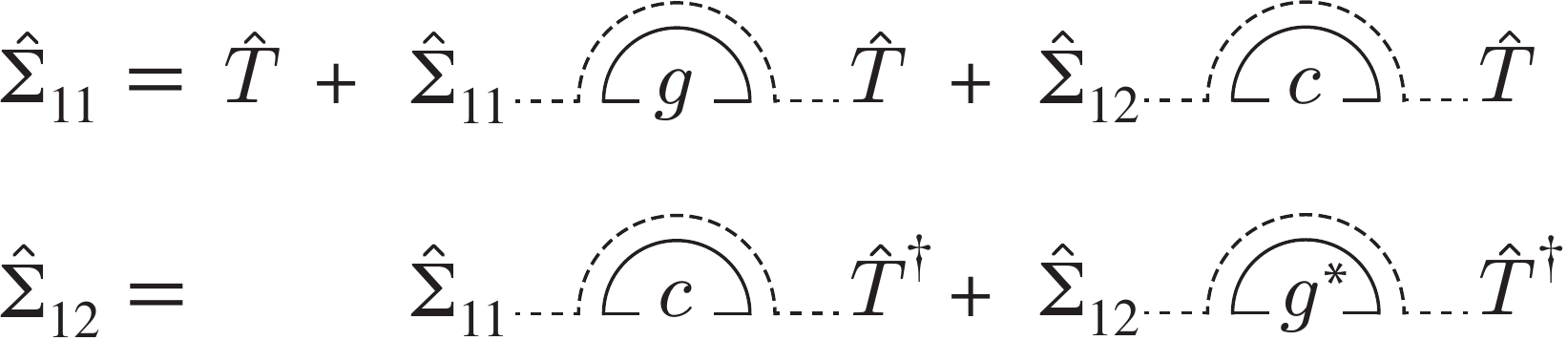}
\caption{
\label{figA4}
%%%%%%%%%%%%%
Coupled equations for the operators $\hat{\Sigma}_{11}$ and $\hat{\Sigma}_{12}$ that define the self-energy $\Sigma = \lim_{\epsilon \to 0^+}\Sigma(Z_{\epsilon})$. Here $g=\lim_{\epsilon \to 0^+ }G_{11}^\epsilon$ and $c=\lim_{\epsilon \to 0^+ }G_{12}^\epsilon$ [see Eq. (\ref{green})].}}
\end{figure}

Equations obeyed by the operators
$\hat{\Sigma}_{11}=\lim_{\epsilon \to 0^+ }\hat{\Sigma}_{11}^\epsilon$ and $\hat{\Sigma}_{12}=\lim_{\epsilon \to 0^+ }\hat{\Sigma}_{12}^\epsilon$
are obtained after summation of all planar rainbow diagrams in the expansion of Fig.\ \ref{figS2} and taking the limit $\epsilon \to 0^+$. The diagrammatic representation of these equations is shown in Fig. \ref{figA4}.
%%%%%FIGA4%%%%%
Equations (\ref{self1}) and (\ref{self2}) of the main text follow after application of `Feynman' rules defined in Fig.\ \ref{figA2}. Furthermore, as follows from Eq.\ (\ref{green}) and the definition of the self-energy matrix, in the limit $\epsilon \to 0^+$, $g$ and $c$ are simply related to  $\Sigma_{11}=\textrm{Tr}\hat{\Sigma}_{11}/N$ and $\Sigma_{12}=\textrm{Tr}\hat{\Sigma}_{12}/N$ by
\begin{equation}
\left[ \begin{array}{cc}
g(z) &c(z)   \\
c(z)& g(z)^*
\end{array} \right]=\left( \begin{array}{cc}
z-\Sigma_{11} & -\Sigma_{12}   \\
-\Sigma_{12}& z^*-\Sigma_{11}^*
\end{array} \right)^{-1}.
\label{LinkGSigma}
\end{equation}
Elimination of the self-energy $\hat{\Sigma}$ from Eqs. (\ref{self1}), (\ref{self2}) and (\ref{LinkGSigma}) yields Eqs. (\ref{sc1}) and (\ref{sc2}) of the main text.

%%%%%%%%%%%%%%   APPENDIX B  %%%%%%%%%%%%%%
\section{Approximate solutions for the borderline of the eigenvalue domain and the eigenvalue density at low density}
\label{appendixB}

Let us show how an explicit equation for the borderline of the support of eigenvalue density of the random Green's matrix (\ref{ERMGreen}) --- Eq.\ (\ref{contour3}) --- can be derived in the low-density limit.
On the one hand, traces appearing in Eqs.\ (\ref{contoura}) and (\ref{contourb}) in the $\ket{\vec{r}}$ representation read
\begin{eqnarray}
\mathrm{Tr} \hat{S} &=&
\mathrm{Tr} \left( \frac{\hat{T}}{1 - g \hat {T}} \right) =
\mathrm{Tr} \left( \hat{T} + g \hat{T} \hat{S} \right)
\nonumber \\
&=& g\iint_{V} \mathrm{d}^3\vec{r}\; \mathrm{d}^3\vec{r'}\, T(\vec{r},\vec{r'})S(\vec{r'},\vec{r}),
\label{contourR1}
\end{eqnarray}
\begin{eqnarray}
\mathrm{Tr} \hat{S} \hat{S}^{\dagger} = \iint_{V}  \mathrm{d}^3\vec{r}\; \mathrm{d}^3\vec{r'} \left|S(\vec{r},\vec{r'})\right|^2,
\label{contourR2}
\end{eqnarray}
where  $T(\vec{r},\vec{r'})=\rho \bra{ \vec{r}}\hat{A}\ket{ \vec{r'}}= \rho \exp(i k_0 | \vec{r}-\vec{r'}|)/k_0 | \vec{r}-\vec{r'}|$ and in Eq.\ (\ref{contourR1}) we used the fact that  $\textrm{Tr} \hat{T} = \rho \textrm{Tr} \hat{A} = 0$, as follows from Eq.\ (\ref{ERMGreen}). On the other hand, $S(\vec{r},\vec{r'})=\bra{ \vec{r}}\hat{S}\ket{ \vec{r'}}$ obeys
\begin{equation}
S(\vec{r},\vec{r'})=T(\vec{r},\vec{r'})+g\int_{V} \mathrm{d}^3\vec{r''} T(\vec{r},\vec{r''})S(\vec{r''},\vec{r'}),
\label{Sinspace}
\end{equation}
as follows from the definition of $\hat{S}$.
Noting that
\begin{equation}
\left(\Delta_{\vec{r}}+k_0^2+i\epsilon\right)T(\vec{r},\vec{r'})
=-\frac{4\pi\rho}{k_0} \delta^{(3)}(\vec{r}-\vec{r}'),
\label{THelmholtz}
\end{equation}
where $\epsilon \to 0^+$, we apply the operator $\Delta_{\vec{r}}+k_0^2+i\epsilon$ to Eq.\ (\ref{Sinspace}) and obtain
\begin{align}
\nonumber
\Delta_{\vec{r}}S(\vec{r},\vec{r'})+
k_0^2\left[1+g\frac{\rho\lambda_0^3}{2\pi^2}\Pi_V(\vec{r})+
i\epsilon\right]S(\vec{r},\vec{r'})&
\\
=-\frac{4\pi\rho}{k_0}\delta^{(3)}(\vec{r}-\vec{r}'),
\label{SHemholtz}
\end{align}
where $\Pi_V(\vec{r})=1$ for $\vec{r} \in V$ and $0$ elsewhere. In the limit of low density $\rho\lambda_0^3 \to 0$, an approximate solution of this equation is obtained by neglecting `reflections' of the `wave' $S(\vec{r},\vec{r'})$ on the boundaries of the volume $V$ and thus setting $\Pi_V(\vec{r}) = 1$ everywhere. This yields $S(\vec{r},\vec{r'})\simeq \rho \exp(i \kappa | \vec{r}-\vec{r'}|)/k_0 | \vec{r}-\vec{r'}|$ with $\kappa(g)=k_0\sqrt{1+g\rho\lambda_0^3/2\pi^2}$.

In order to evaluate the integrals (\ref{contourR1}) and (\ref{contourR2}), we will make use of the following auxiliary result:
\begin{equation}
\iint_{V(R)} \frac{\mathrm{d}^3\vec{r}}{V}\frac{\mathrm{d}^3\vec{r'}}{V}
f(\vert\vec{r}-\vec{r'}\vert)=24\int_0^1 \mathrm{d} x f(2Rx) s(x) x^2,
\label{sphericalInt}
\end{equation}
where $f$ is an arbitrary function, $V(R)=4\pi R^3/3$, and $s(x)=1-3x/2+x^3/2$. To derive this equation, we define new variables $\vec{x}=(\vec{r}-\vec{r'})/2R$ and $\vec{y}=(\vec{r}+\vec{r'})/2R$. The conditions $r\leq R, r'\leq R$ become $x^2+y^2+2xyt\leq 1$, with $0\leq t \leq 1$, so that
\begin{equation}
\iint_{V(R)}\!\! \frac{\mathrm{d}^3\vec{r}}{V}\frac{\mathrm{d}^3\vec{r'}}{V}(...)
=\frac{18}{\pi}\int_{V(1)}\!\!\!\!\!\!
\mathrm{d}^3 \vec{x}\int_0^1\!\!\! \mathrm{d}t \!\int_0^{y_M(t,x)}\!\!\!\!\!\!\!\!\!\!\!
\mathrm{d}y\,y^2(...),
\label{sphericalInt2}
\end{equation}
where  $y_M(t,x)=\sqrt{1+(t^2-1)x^2}-tx$. Evaluation of all integrals except one in Eq.\ (\ref{sphericalInt2}) leads to Eq.\ (\ref{sphericalInt}).

We now plug the explicit expressions for $T(\vec{r},\vec{r'})$ and $S(\vec{r},\vec{r'})$ into Eqs.\ (\ref{contourR1}) and (\ref{contourR2}) and use Eq.\ (\ref{sphericalInt}). This yields
\begin{align}
&\mathrm{Tr} \hat{S} = 2\gamma N g h[-i\kappa(g)R-ik_0R],&
\label{contourR3}
\\
&\mathrm{Tr} \hat{S} \hat{S}^{\dagger} = 2\gamma N h[2\textrm{Im} \kappa(g)R],
\label{contourR4}
\end{align}
with
\begin{align}
h(x)&=\frac{\int_0^1 \mathrm{d}u s(u)e^{-2ux}}{\int_0^1 \mathrm{d} u s(u)}&
\nonumber
\\
&=\frac{1}{6x^4}[3-6x^2+8x^3-3(1+2x)e^{-2x}],
\label{defh}
\end{align}
and $\gamma = 9N/8(k_0R)^2$. In the low density limit, the latter is equal to the  second moment of the absolute value of $\Lambda$: $\gamma = \langle |\Lambda|^2 \rangle$. We checked numerically that even at higher densities (at least, up to $\rho \lambda_0^3 \sim 100$), $\gamma$ is still a good approximation for $\langle |\Lambda|^2 \rangle$ and hence a meaningful parameter.

In the low-density limit, $g$ can be eliminated from Eqs.\ (\ref{contoura}) and (\ref{contourb}) by neglecting $\mathrm{Tr} \hat{S}/N$ in Eq.\ (\ref{contoura}) and substituting $g = 1/z$ into Eq.\ (\ref{contourR4}). This yields Eq.\ (\ref{contour2}) and then Eq.\ (\ref{contour3}), if the argument of the function $h$ in Eq.\ (\ref{contour2}) is expanded in series in $\rho \lambda_0^3$. By comparing Eq.\ (\ref{contour3}) with the exact solution obtained in Appendix \ref{appendixC}, we conclude that it is valid up to densities as high as $\rho \lambda_0^3 \simeq 10$.

Finally, Eqs.\ (\ref{contourR3}) and (\ref{contourR4}) for $\mathrm{Tr} \hat{S}$ and $\mathrm{Tr} \hat{S} \hat{S}^{\dagger}$ can be used to find the resolvent $g(z)$ using Eq.\ (\ref{resolvent}) and then the density of eigenvalues $p(\Lambda)$ using Eq.\ (\ref{pnonherm}).

%%%%%%%%%%%%%%   APPENDIX C  %%%%%%%%%%%%%%
\section{Exact solution for the borderline of the eigenvalue domain at any density}
\label{appendixC}

In this Appendix we show how Eqs.\ (\ref{contoura}) and (\ref{contourb}) can be solved exactly using the bi-orthogonal basis of right $\ket{\mathcal{R}_{\alpha}}$ and left $\ket{\mathcal{L}_{\alpha}}$ eigenvectors of $\hat{T}$. These eigenvectors obey $\hat{T}\ket{\mathcal{R}_{\alpha}}=\mu_{\alpha}\ket{\mathcal{R}_{\alpha}}$ and $\hat{T}^\dagger\ket{\mathcal{L}_{\alpha}}=\mu_{\alpha}^*\ket{\mathcal{L}_{\alpha}}$. In this basis, Eqs.\ (\ref{contoura}) and (\ref{contourb}) read
\begin{align}
&z = \frac{1}{g} + \frac{g}{N} \sum_{\alpha}\frac{\mu_\alpha^2}{1-g\mu_\alpha},&
\label{contourT1}
\\
&\frac{1}{|g|^2} = \frac{1}{N} \sum_{\alpha,\beta}
\frac{\mu_\alpha \mu_\beta^*
\ps{\mathcal{L}_\alpha}{\mathcal{L}_\beta}
\ps{\mathcal{R}_\beta}{\mathcal{R}_\alpha}
}
{(1-g\mu_\alpha)(1-g\mu_\beta)^*},
\label{contourT2}
\end{align}
where, similarly to the derivation in Appendix \ref{appendixB}, we made use of the fact that $\mathrm{Tr} \hat{T} = 0$ and therefore $\mathrm{Tr} \hat{S} = g \mathrm{Tr} \hat{T} \hat{S}$.
The problem essentially reduces to solving the eigenvalue equation
\begin{equation}
\rho\int_V \mathrm{d}^3 \vec{r'} \frac{\exp(i k_0 | \vec{r}-\vec{r'}|)}{k_0 | \vec{r}-\vec{r'}|}\mathcal{R}_{\alpha}(\vec{r'})=
\mu_\alpha\mathcal{R}_{\alpha}(\vec{r}),
\label{rightreal}
\end{equation}
where $\vec{r} \in V$. As follows from Eq.\ (\ref{THelmholtz}), $\mathcal{R}_{\alpha}(\vec{r})$ is also an eigenvector of the Laplacian operator, $\Delta_{\vec{r}}\mathcal{R}_{\alpha}(\vec{r})=-\kappa_{\alpha}^2\mathcal{R}_{\alpha}(\vec{r})$, with $\kappa_{\alpha}=\kappa(1/\mu_\alpha)$. In a sphere of radius $R$, using the decomposition of the kernel of Eq.\ (\ref{rightreal}) in spherical harmonics, it is quite easy to find that \cite{svid10}
\begin{equation}
\mathcal{R}_{\alpha}(\vec{r}) = \mathcal{R}_{lmp}(\vec{r})=\mathcal{A}_{lp}j_l(\kappa_{lp}r)Y_{lm}(\theta,\phi),
\end{equation}
where $\theta$ and $\phi$ are the polar and azimuthal angles of the vector $\vec{r}$, respectively, $j_l$ are spherical Bessel functions of the first kind, $Y_{lm}$ are spherical harmonics, $\mathcal{A}_{lp}$ are normalization coefficients, and $\alpha=\{ l,m,p \}$. Furthermore, coefficients $\kappa_{lp}$ obey \cite{svid10}
\begin{equation}
\frac{\kappa_{lp}}{k_0}=\frac{j_l(\kappa_{lp}R)}{j_{l-1}(\kappa_{lp}R)}\frac{h^{(1)}_{l-1}(k_0R)}{h^{(1)}_l(k_0R)},
\label{modesk}
\end{equation}
where  $h^{(1)}_l$ are spherical Hankel functions. Integer $p$ labels the different solutions of this equation for a given $l$. Hence, eigenvalues $\mu_{lp}=\rho\lambda_0^3/2\pi^2(\kappa_{lp}^2/k_0^2-1)$ are $(2l+1)$-times degenerate ($m\in [-l,l]$).

In the limit $k_0R \gg 1$, we can use asymptotic expressions for the spherical functions in Eq.\ (\ref{modesk}) to obtain
\begin{equation}
\frac{i}{2}\textrm{ln}\left(\frac{\kappa_{lp}+k_0}{\kappa_{lp}-k_0}\right)=-\kappa_{lp}R+\left(\frac{l}{2}+p\right)\pi.
\end{equation}
In this limit, the eigenvalues $\mu_{lp}$ are therefore localized in the vicinity of a roughly circular line in the complex plane given by
\begin{equation}
\left|\frac{\kappa(1/\mu)-k_0}{\kappa(1/\mu)+k_0} \right|^2  \left|e^{4i\kappa(1/\mu)R} \right|=1.
\label{largerho}
\end{equation}

Let us now study the eigenvectors. Using standard properties of spherical harmonics and spherical Bessel functions \cite{morse}, we can show that
\begin{align}
&\ps{\mathcal{R}^*_{lmp}}{\mathcal{R}_{l'm'p'}}
=(-1)^m\mathcal{A}_{lp}^2\frac{R^3}{2}\Big[j_l(\kappa_{lp}R)^2&
\nonumber
\\
&-j_{l-1}(\kappa_{lp}R)j_{l+1}(\kappa_{lp}R)\Big]
\delta_{l,l'}\delta_{m,-m'}\delta_{p,p'}.
\end{align}
From the normalization condition $\ps{\mathcal{L}_{lmp}}{\mathcal{R}_{l'm'p'}}
=\delta_{l,l'}\delta_{m,m'}\delta_{p,p'}$, we find that $\mathcal{L}_{lmp}(\vec{r})=(-1)^m\mathcal{R}_{l(-m)p}(\vec{r})^*$ and
\begin{equation}
\mathcal{A}_{lp}=\sqrt{\frac{2}{R^3}}\frac{1}{\sqrt{j_l(\kappa_{lp}R)^2-j_{l-1}(\kappa_{lp}R)j_{l+1}(\kappa_{lp}R)}}.
\end{equation}
On the other hand, we also have
\begin{align}
&\ps{\mathcal{R}_{lmp}}{\mathcal{R}_{l'm'p'}}=
\frac{R^2\mathcal{A}^*_{lp}\mathcal{A}_{lp'}}{\kappa_{lp'}^2-
\kappa_{lp}^{*2}}\Big[\kappa_{lp}^* j_{l-1}(\kappa_{lp}^*R) j_{l}(\kappa_{lp'}R)&
\nonumber
\\
&-\kappa_{lp'} j_{l-1}(\kappa_{lp'}R) j_{l}(\kappa_{lp}^*R)\Big]\delta_{l,l'}\delta_{m,m'},
\end{align}
and $\ps{\mathcal{L}_{lmp}}{\mathcal{L}_{l'm'p'}}=\ps{\mathcal{R}_{lmp}}{\mathcal{R}_{lmp'}}\delta_{l,l'}\delta_{m,m'}$. It is now convenient to introduce a new coefficient
\begin{widetext}
\begin{equation}
C_{lpp'}=
\frac{4
\bigg[\kappa_{lp}^*R j_{l-1}(\kappa_{lp}^*R) j_{l}(\kappa_{lp'}R)-\kappa_{lp'}R j_{l-1}(\kappa_{lp'}R) j_{l}(\kappa_{lp}^*R)\bigg]^2
}
{
\bigg[\kappa_{lp'}^2R^2-\kappa_{lp}^{*2}R^2\bigg]^2
\bigg[j_l(\kappa^*_{lp}R)^2-j_{l-1}(\kappa^*_{lp}R)j_{l+1}(\kappa^*_{lp}R)\bigg]
\bigg[j_l(\kappa_{lp'}R)^2-j_{l-1}(\kappa_{lp'}R)j_{l+1}(\kappa_{lp'}R)\bigg]
},
\label{coeffC}
\end{equation}
\end{widetext}
in terms of which Eqs. (\ref{contourT1}) and (\ref{contourT2}) become
\begin{align}
&z =\frac{1}{g}+\frac{g}{N} \sum_l \sum_p\frac{(2l+1)\mu_{lp}^2}{1-g\mu_{lp}},&
\label{contourT3}
\\
&\frac{1}{|g|^2} = \frac{1}{N}\sum_l \sum_p\sum_{p'}
\frac{(2l+1)\mu_{lp'}\mu_{lp}^*C_{lpp'}}
{(1-g\mu_{lp'})(1-g\mu_{lp})^*}.
\label{contourT4}
\end{align}

To find the borderline of the support of eigenvalue density of the matrix (\ref{ERMGreen}) shown in Figs.\ \ref{fig1}(c) and (d), we apply the following recipe. (1) Find solutions $\kappa_{lp}$ of Eq.\ (\ref{modesk}) numerically and then compute the corresponding $\mu_{lp}$.  (2) Compute the coefficients $C_{lpp'}$ using Eq.\ (\ref{coeffC}). (3) Find lines on the complex plane $g$ defined by Eq.\ (\ref{contourT4}). (4) Transform the lines on the complex plane $g$ into contours on the complex plane $z$ using Eq.\ (\ref{contourT3}). The latter contours are the borderlines of the support of eigenvalue density $p(\Lambda)$.

%=====================================================================================

\end{document}